\documentclass[11pt,a4paper]{article}

\usepackage[utf8]{inputenc}
\usepackage[T1]{fontenc}
\usepackage[margin=1in]{geometry}
\usepackage{amsmath, amssymb}
\usepackage{booktabs}
\usepackage{array}
\usepackage{graphicx}
\usepackage{xcolor}
\usepackage{listings}

\usepackage{caption}
\usepackage{enumitem}
\usepackage{titlesec}
\usepackage{float}
\usepackage{hyperref}

\usepackage[
  backend=biber,
  style=numeric,
  sorting=nyt,
  doi=true,
  url=true,
  eprint=true,
  maxbibnames=99
]{biblatex}

\hypersetup{
  colorlinks=true,
  linkcolor=black,
  urlcolor=blue!60!black,
  citecolor=blue!50!black,
  pdftitle={AgensFlow: A Coordination-Policy Substrate for Multi-Agent Systems},
  pdfauthor={Nicole Koenigstein},
}
\titleformat{\section}{\Large\bfseries}{\thesection}{1em}{}
\titleformat{\subsection}{\large\bfseries}{\thesubsection}{1em}{}

\title{
  \vspace{-2em}
  \huge\bfseries AgensFlow: A Coordination-Policy Substrate for Multi-Agent Systems
}
\author{Nicole Koenigstein\thanks{Independent researcher.
Repository: \url{https://github.com/Nicolepcx/AgensFlow}.
Correspondence: \texttt{nicole@nicolekoenigstein.com}}}
\date{\today}

\begin{document}
\maketitle

\begin{abstract}
\noindent
Multi-agent systems built on large language models (LLMs) require many coordination choices that are difficult to fix a priori: which skill protocol to invoke, which agent role should perform a subtask, which model to bind to each role, how roles should interact, when to use retrieval or verification, and when to omit a step entirely. These choices interact with task regime and operational constraints, so static pipelines and one-off model comparisons provide only a limited view of the design space. This paper introduces AgensFlow, an open-source framework that treats multi-agent coordination as an online policy-learning problem under partial observability. The framework makes
coordination decisions observable and learnable from repeated trajectories, rather than treating skill, role, model, topology, and evaluation choices as fixed pipeline design.

AgensFlow is evaluated on two corpora: distributed-systems incident tasks and security-advisory tasks. The evaluation shows three main results: learned routing reaches a higher-quality operating point than a fixed pipeline baseline on coordination-heavy classes; \texttt{skip:X} isolates topology compression as a meaningful part of the substrate; and warm-started policy graphs can reduce exploration cost while preserving plateau quality. Overall, the results support that learned, auditable routing can improve coordination-heavy multi-agent workflows over static wiring.
\end{abstract}

\section{Introduction}
\label{sec:intro}

AI agents \cite{Wang_2024} are language-model-driven systems that operate through iterative cycles of planning, action, tool use, and feedback. Moving beyond isolated prompting settings, these systems now function in environments where they interact with tools, external resources, and complex organizational workflows. As their capabilities scale, agents are being evaluated and deployed across diverse applied domains, ranging from software engineering \cite{yang2024sweagentagentcomputerinterfacesenable, tang2024codeagentautonomouscommunicativeagents}
and medical support \cite{kim2024healthllmlargelanguagemodels, wang2026medmemorybenchbenchmarkingagentmemory}
to financial analysis \cite{bigeard2025financeagentbenchmarkbenchmarking, okpala2025agenticaisystemsapplied}
and scientific discovery \cite{yamada2025aiscientistv2workshoplevelautomated, mitchener2025kosmosaiscientistautonomous}. Furthermore, they are increasingly studied as embedded digital workers capable of navigating internal company infrastructure and collaborating with human colleagues \cite{xu2025theagentcompanybenchmarkingllmagents}.

As agentic workloads become more complex, longer-horizon, and more dependent on heterogeneous capabilities, a single-agent framing becomes insufficient. Multi-agent systems (MAS) \cite{wu2023autogen,li2023camel} have therefore gained attention as a way to distribute work across specialized roles, support parallel
exploration, introduce verification steps, and coordinate task decomposition.

\begin{figure}[H]
  \centering
  \includegraphics[width=\linewidth]{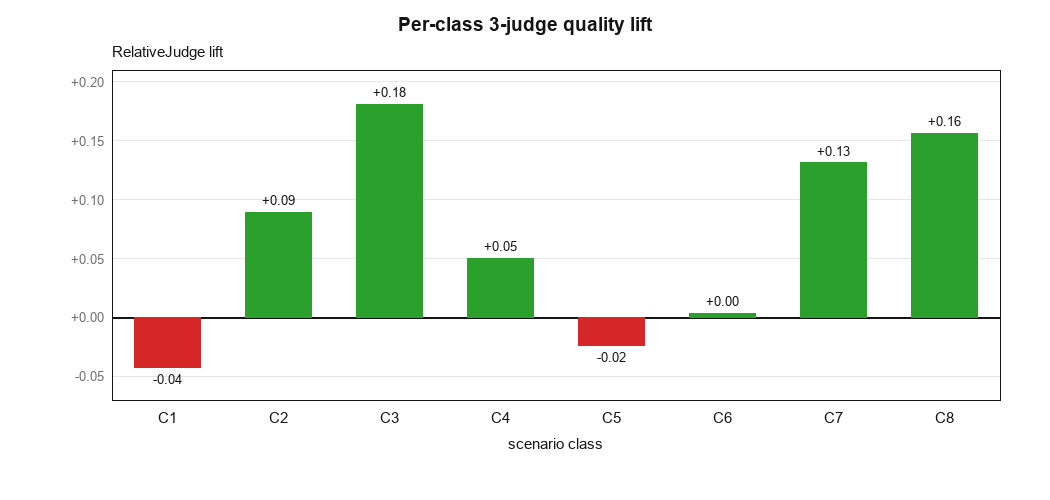}
  \caption{\textbf{Per-class quality lift under 3-judge audit.}
    Learned routing improves most strongly on coordination-heavy
    classes, especially C3 cross-document multi-vendor reasoning, C7
    mitigation correctness, and C8 cross-vendor pair tasks. Procedural
    classes are ties or narrow trades.}
  \label{fig:e09-per-class-lift}
\end{figure}

This move toward specialization also increases the need for explicit task guidance. Agent skills provide structured procedural knowledge that augments agents at inference time \cite{ling2026agentskillsdatadrivenanalysis}. Recent work shows that curated skills can improve task performance, although their effects vary substantially across domains and tasks \cite{li2026skillsbenchbenchmarkingagentskills}. Skills therefore function not only as procedural support, but also as a coordination-relevant design choice: the system must decide which skill, if any, is appropriate for a given task context. Figure~\ref{fig:e09-per-class-lift} previews why this shift matters empirically: learned routing improves most on coordination-heavy task classes while trading narrowly on procedural or out-of-corpus cases.

Taken together, these developments shift the central technical bottleneck from isolated agent capabilities to dynamic coordination. A robust multi-agent system can no longer rely on static, hardcoded pipelines; it must actively decide which skills to invoke, which models to bind to specific roles, what coordination topology \cite{kim2026sciencescalingagentsystems} to deploy, and when a costly retrieval or verification step can be skipped entirely. These choices do not exist in isolation, but rather lie on a joint design surface spanning at least four interacting axes:

\begin{itemize}[leftmargin=*, itemsep=0.3em, topsep=0.3em]
  \item \textbf{Task signature.} The type of task being encountered
    (i.e. single-document factual, cross-document multi-source synthesis,
    ambiguous out-of-corpus, coordination-heavy) and the structural
    features that distinguish it.
  \item \textbf{Skill protocol.} The procedural constraint or guidance
    supplied to the agent, such as concise answering, evidence-citation
    discipline, retrieval-grounded synthesis, verification, or domain-
    specific task handling.
  \item \textbf{Model binding.} The underlying model assigned
    to a given role or skill, where each option occupies a different
    capability, cost, and latency point.
  \item \textbf{Coordination topology.} Which roles, skills, or
    verification cells run, in what order, with which dependencies, and
    which steps can be \emph{omitted entirely} on a per-task basis.
\end{itemize}

These axes interact. A configuration that works well for one task signature may fail under another, and the effects are not additive: changing the model, skill protocol, topology, or verification cadence can alter the behaviour of the entire system. The design problem is therefore not to choose "the best model", but to learn an operating policy over a large and shifting coordination surface. This surface cannot be explored reliably through intuition or isolated benchmark comparisons. Fixed benchmarks typically evaluate only a small set of configurations, making it difficult to observe how coordination choices interact across repeated task trajectories. This motivates a coordination layer that can observe repeated trajectories, update routing priors, and make the skill, model, topology, and evaluation choices behind a multi-agent system inspectable.

To address this gap, this paper introduces AgensFlow, an open-source coordination-policy substrate for multi-agent systems. The name combines Latin \emph{agens}, meaning acting, driving, or conducting, with \emph{flow}: the framework is concerned not with a single static agent, but with agency in motion, structured through reusable coordination decisions. AgensFlow makes use-case-conditioned skill selection, model-role assignment, topology choice, and reward audit observable and learnable from repeated trajectories, rather than fixing them as one-off pipeline decisions. The framework contributes three elements: (i) a formulation of multi-agent orchestration as online policy learning under partial observability, (ii) an inspectable policy graph over skills, models, and topology actions, and (iii) reward-signal auditability as a first-class part of system design. The empirical sections evaluate this substrate through cross-domain validation, no-skip ablation, warm-start transfer, and cross-judge auditing.

\section{Related Work}
\label{sec:related}

\subsection{Reasoning and Agent Design Patterns}
Chain-of-thought prompting \cite{wei2022cot} introduced an influential way to elicit intermediate reasoning traces from language models, while ReAct-style methods \cite{yao2023react} connected such reasoning traces to tool and environment actions. More recent reasoning-oriented models make this capability more explicit by allocating additional inference-time computation to intermediate reasoning before producing an answer. This can improve tasks that require multi-step inference, decomposition, or tool-mediated problem solving, but it is not uniformly beneficial. For simpler tasks, or tasks where concise retrieval-grounded synthesis is sufficient, extended reasoning can increase token usage, latency, cost, and the number of possible failure paths.

In AgensFlow, reasoning patterns are therefore not treated as the definition of an agent, nor as universally preferable execution modes. They are represented as \emph{skill cards}: structured behavioural constraints within the variant pool that can be selected, combined, or skipped depending on the task signature. The framework's contribution is to learn which reasoning constraint, model binding, and coordination topology should be selected for a given task class, rather than assuming that more explicit reasoning is always the correct choice.

\subsection{Conversation-Based Multi-Agent Systems}

AutoGen \cite{wu2023autogen} and CAMEL \cite{li2023camel} are prominent examples of conversation-based MAS: agents communicate through free-form natural-language transcripts, and coordination is mediated through role conditioned dialogue. This design is expressive, but it introduces a structural trade-off: coordination decisions become entangled with transcript content and are difficult to fold into a reusable, inspectable representation. AgensFlow takes a different approach. Coordination is encoded as \emph{structured handoffs} over a typed schema, which allows the policy graph to fold experience across runs and makes per-signature value estimates auditable.

\subsection{Learned Tool Use and Orchestration Policies}

Another related line studies how agents learn to select tools, interfaces, or execution strategies rather than relying entirely on handwritten tool-use policies. Toolformer \cite{schick2023toolformer} showed that language models can be trained to decide when API calls are useful, and subsequent agent systems have increasingly treated tool choice, memory access, and execution control as learnable or measurable parts of the system rather than fixed prompting patterns. Recent work on scaling agent systems \cite{kim2026sciencescalingagentsystems} similarly argues that agent performance depends on system-level coordination choices, not only on the base model.

AgensFlow is closest to this line of work, but differs in the object being learned. The policy is not only a tool selector or planner switch. It is a persistent, auditable routing policy over use-case signatures, with a joint action surface spanning skill protocols, model-role bindings, optional topology cells, and termination. The \texttt{skip:X} action makes omission itself a first-class topology decision, and the reward layer is exposed to cross-judge audit rather than treated as a fixed evaluator.

\subsection{Modular Expert Coordination}

A separate line of work studies systems composed of specialised modules whose activation is selected conditionally on context. Recurrent Independent Mechanisms (RIM) \cite{goyal2021recurrent}, for example, formalise sparse context-dependent module activation for sequence models, supporting compositional behaviour and improved generalisation under changing conditions.

AgensFlow builds on this structural view of intelligence as context-dependent expert coordination. The author's prior work \cite{koenigstein2023dynamic} applied this idea to complex, non-stationary financial time-series prediction, where expert modules were selected through attention conditioned on market state and news sentiment. AgensFlow carries the same core intuition into multi-agent systems: specialised capabilities should not be wired statically, but selected according to the current task regime and accumulated evidence. The framework changes the substrate from differentiable attention over time-series experts to an inspectable online policy over agent skills, model-role bindings, and topology choices.

\subsection{Relative Trajectory Evaluation}

Agentic systems are difficult to score from final answers alone, because two trajectories can reach plausible outputs through very different evidence, tool-use, and verification paths. Relative trajectory evaluation addresses this by comparing multiple rollouts for the same task side-by-side against an explicit rubric, rather than asking a judge to assign an isolated absolute score. This makes the evaluation signal more sensitive to coordination quality, recovery behaviour, and evidence use, and it gives the learning substrate a reward signal closer to the routing decisions it is trying to improve. AgensFlow's \texttt{RelativeJudge} follows this general idea and is inspired by RULER-style relative LLM evaluation \cite{ruler2025}; its concrete integration with the framework is described in \S\ref{sec:reward-audit}.

\section{Preliminaries: Coordination as Policy Learning}
\label{sec:problem}

AgensFlow treats multi-agent orchestration as a partially observable sequential decision process and learns a policy over an abstracted state space. The system does not observe the user's full intent, latent task difficulty, evidence quality, model reliability, or intermediate reasoning quality directly. Instead, it constructs an observable folded signature $s_t$ from typed task features, structured handoff state, and belief estimates. At each routing step, the system observes $s_t$, chooses one legal coordination action $a_t$, observes the updated handoff state after that action, and eventually backs up the trajectory-level reward to the visited $(s_t,a_t)$ pairs. The policy is therefore learned over reusable abstract states rather than over individual prompts. In this sense, the folded signature is a state abstraction \cite{li2006stateabstraction}: it trades fine-grained state fidelity for value sharing across related runs.

\subsection{State as Folded Task Signatures}

The true state of an agentic task includes the user's intent, the latent difficulty of the question, the quality and coverage of the retrieved evidence, the hidden capabilities and failure modes of each model, and the intermediate reasoning quality of each agent. This state is not directly observable. AgensFlow therefore operates on an observable, belief-conditioned signature:

\begin{equation}
s_t = \phi(x_t, h_t, b_t)
    = (\ell_t, z_t, \beta(\hat c_t), \beta(\hat u_t),
       \beta(\hat r_t), \beta(\hat e_t)).
\end{equation}

where $x_t$ denotes the task context, $h_t$ denotes the structured handoff state, $b_t$ denotes the current belief vector, and $\phi$ maps these observations to a discrete signature. The regime label $\ell_t$ is produced by the default rule-based regime detector from typed task features: ambiguity level, contradiction risk, evidence availability, and verification need. The supported regime labels are \texttt{straightforward}, \texttt{evidence\_heavy}, \texttt{ambiguous}, \texttt{contradictory}, \texttt{high\_risk}, and \texttt{exploratory}. 

$z_t \in \{0,1\}^7$ is the binary handoff mask for whether \texttt{goal}, \texttt{subproblem}, \texttt{evidence}, \texttt{critique}, \texttt{verification}, \texttt{draft\_answer}, and \texttt{merged\_answer} are populated; and $\beta(\cdot)$ are continuous belief estimates to the configured signature granularity. The four signature-folded belief terms are estimated correctness $\hat c_t$, estimated uncertainty $\hat u_t$, estimated contradiction risk $\hat r_t$, and estimated evidence sufficiency $\hat e_t$. The runtime additionally tracks an \emph{estimated handoff quality} belief that is updated alongside the other four but is intentionally excluded from the signature, to keep the policy graph compact; it is available for inspection in traces but does not influence $\phi$. These belief estimates are heuristic in the current release. They are updated from observed agent contributions across six agent roles: the planner improves handoff quality and modestly reduces uncertainty when a subproblem is set; memory increases evidence sufficiency in proportion to the retrieved evidence count and also lowers uncertainty and lifts handoff quality; the solver, when a draft answer is produced, raises correctness, reduces uncertainty, and lifts handoff quality; the critic raises contradiction risk and slightly raises uncertainty whenever a critique is produced; the verifier parses its verdict and updates correctness, uncertainty, contradiction risk, and (under a supported verdict) evidence sufficiency; and the synthesiser, when a merged answer is produced, modestly raises correctness and handoff quality and lowers uncertainty. The evaluator does not produce belief deltas; its decision feeds the reward function rather than the belief state. 

The granularity of $\beta$ controls the abstraction's bias-variance tradeoff. Coarser bins create fewer signatures and more value sharing, but risk aliasing task states that need different routing policies. Finer bins preserve more distinctions, but require more data before each signature becomes reliable. Two trajectories are said to fold to the same signature when their observations map to the same $s_t$ under $\phi$, allowing value estimates to be shared in the policy graph. This signature is the central compression that makes online learning feasible: the substrate does not memorise individual prompts; it learns reusable coordination behaviour for recurring task regimes. The cross-domain experiment in \S\ref{subsec:cross-domain} tests this reuse across new prompts within the same scenario classes.

\subsection{Actions: Skills, Models, and Topology}

At each routing step the policy chooses one action, executes it, observes the resulting handoff update, recomputes the signature, and then chooses again. The policy therefore does not commit to a full trajectory upfront. It performs sequential, myopic action selection with delayed trajectory-level reward backup. The available action set is state-dependent:

\begin{equation}
  a_t \in \mathcal{A}(s_t)
    = \{\text{invoke}(k,m)\}
      \cup \{\texttt{skip}:X\}
      \cup \{\text{terminate}\}
\end{equation}

where $(k,m)$ denotes a skill protocol bound to a model and \texttt{skip}:$X$ denotes omitting a still-scheduled cell from the current trajectory. The legal set $\mathcal{A}(s_t)$ is determined by the activation plan, completed handoff fields, budget state, and termination rules. \texttt{skip}:$X$ is admitted into the candidate set for any $X$ that is still scheduled in the plan and not yet invoked, provided at least one other legal action remains so that the run can still finish; the framework does not partition skills into "required" and "optional" categories. \texttt{terminate} is included in the formal action set for completeness, but in the current runtime it is not a policy choice the router selects. It is triggered implicitly when one of four conditions becomes true: the evaluator marks the run complete, the per-run budget is exhausted, no legal actions remain, or the governance layer halts the run on a policy violation. The policy can therefore cause termination by choosing to invoke the evaluator, but never picks
\texttt{terminate} directly. This is the key distinction from a fixed pipeline: topology is not merely configuration, but part of the action space. The policy can learn that one regime benefits from memory plus verification, while another should bypass retrieval or skip a redundant solver entirely.

\subsection{Reward: Quality Under Operational Constraints}

Trajectory-level reward is observed only after the run completes, and the quality component is stochastic because it is estimated by an LLM judge. AgensFlow therefore composes judged trajectory quality with operational penalties:

\begin{equation}
  r(t) = w_q Q(t) - w_c C(t) - w_\rho \rho(t)
\label{eq:hybrid-reward}
\end{equation}

where $Q(t)$ is the \texttt{RelativeJudge} quality score for trajectory $t$, $C(t)$ is normalized token cost, and $\rho(t)$ is a retry or failure penalty. $Q(t) \in [0,1]$ is produced by same-task relative trajectory scoring and then reduced to one scalar per trajectory; \S\ref{sec:reward-audit} describes the judge protocol and cross-judge audit. In the reported experiments, the default weights are $w_q=1.0$, $w_c=0.3$, and $w_\rho=0.15$, with token cost normalized by an 8,000-token cap. The learning problem is therefore not to minimize cost or maximize judge score in isolation, but to find stable coordination policies that improve judged task performance while keeping operational cost and failure modes visible.

\subsection{Learned Object: An Auditable Policy Graph}

The learned object is a policy graph keyed by folded signature. For each $(s,a)$ pair, the graph stores visits, mean reward, reward variance, token statistics, and failure counts. Action selection uses a reliability-aware UCB1 variant \cite{auer2002ucb} with annealed exploration and an explicit failure penalty:

\begin{equation}
  \mathrm{score}(s,a) =
  \bar r(s,a)
  + c_s \sqrt{\frac{\log(N_s + 1)}{N_{s,a}}}
  - \lambda f(s,a),
\label{eq:ucb-reliability}
\end{equation}

where $\bar r(s,a)$ is the mean backed-up reward, $N_s$ is the signature visit count, $N_{s,a}$ is the action visit count, and $f(s,a)$ is the recorded failure rate for that edge. The exploration coefficient is annealed per signature as

\begin{equation}
  c_s = \max(0.5, 1.4 \cdot 2^{-N_s/50}),
\end{equation}

where $1.4$ is the initial UCB1 exploration constant, $50$ is the visit half-life, and $0.5$ is the minimum exploration floor. Equivalently, the exploration bonus is multiplied by $0.5^{N_s/50}$ as a signature accumulates visits: at $N_s=0$, $c_s=1.4$; at $N_s=50$, $c_s=0.7$; after roughly 75 visits the floor activates and $c_s$ remains $0.5$. The intent is to explore widely when a folded signature is new, narrow exploration as repeated reward observations accumulate, and avoid collapsing to pure exploitation permanently.

The default reliability weight is $\lambda=0.5$. Truly unvisited actions receive infinite score to force initial exploration. Failure counts are not only logged for inspection: they enter the acquisition rule through $-\lambda f(s,a)$, downweighting actions that repeatedly trigger validation or recoverable execution failures even when their final reward is acceptable. Validation failures are schema or contract-check rejections during typed agent I/O; recoverable execution failures are failed attempts that later succeed through retry or correction. The graph is persistent and auditable: after a run, an operator can inspect which actions the system learned to prefer for which task regimes.

\section{Method: The AgensFlow Coordination Substrate}
\label{sec:method}

\begin{figure}[H]
  \centering
  \includegraphics[width=\linewidth]{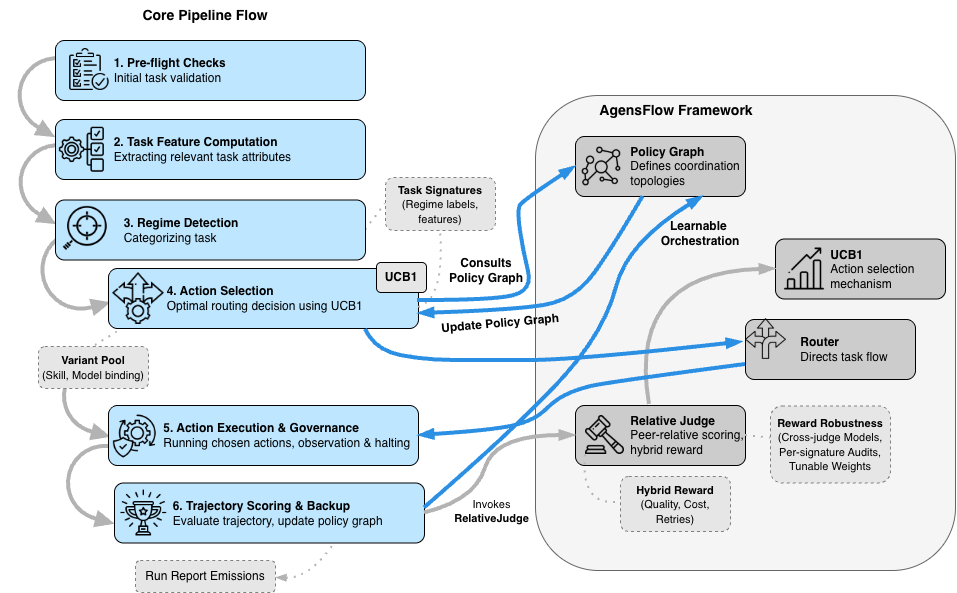}
  \caption{\textbf{AgensFlow at a glance.} Left: the per-task
    lifecycle (steps 1 to 6), from pre-flight gating through task
    feature computation, regime detection, UCB1-based action
    selection over the variant pool, execution, and trajectory
    scoring with policy-graph backup. Right: the framework's
    persistent components and how they interact with the
    lifecycle: the policy graph that UCB1 consults, the router
    that dispatches actions, \texttt{RelativeJudge} with
    its hybrid reward and audit hooks. Note that the
    \emph{six steps} on the left are a procedural per-task flow;
    they should not be confused with the \emph{six dependency
    layers} of the framework (Table~\ref{tab:layers} below), which
    are a structural decomposition where every layer is active
    simultaneously at runtime.}
  \label{fig:overview}
\end{figure}

Figure~\ref{fig:overview} summarizes the runtime lifecycle and the
persistent substrate components before the individual design principles
are unpacked below.

\subsection{Design Principles}
\label{subsec:design-principles}

AgensFlow implements the formulation above as a coordination-policy substrate. The method is built around four ideas:

\begin{enumerate}[leftmargin=*, itemsep=0.3em, topsep=0.3em]
  \item \textbf{Task signatures.} Every incoming task is folded
    into the signature defined in \S\ref{sec:problem}: a regime
    label, a binary handoff mask, and bucketed belief estimates.
    Structural task features such as ambiguity, contradiction risk,
    evidence availability, and verification need feed the regime
    detector; they are not separate signature coordinates unless
    encoded through the regime label or handoff state. Tasks that
    fold to the same signature share learning.
  \item \textbf{Variant pool.} The router's action space at every
    step exposes combinations of \emph{skill protocol} (e.g.\
    \texttt{solver\_concise}, \texttt{solver\_cot},
    \texttt{solver\_evidence}) $\times$ \emph{model binding}
    (\texttt{haiku} / \texttt{fast} / \texttt{mini}), concretely
    nine solver variants in the evaluated skill-variant configuration are used
    throughout the experiments in this paper.
  \item \textbf{\texttt{skip:X} topology learning.} The action
    space at each step also includes \texttt{skip:X} for any
    optional cell $X$. The policy can learn to \emph{exclude}
    skills from a trajectory, not only to re-order or re-bind
    them. This is the mechanism that converts a static pipeline
    decision into a per-task-class learnable decision.
  \item \textbf{Reliability-aware UCB1 on folded signatures.} The
    acquisition rule in Equation~\ref{eq:ucb-reliability} extends
    UCB1 \cite{auer2002ucb} with annealed exploration and a
    per-edge failure penalty. The substrate therefore trades off
    mean reward, exploration, and observed reliability when choosing
    the next legal action.
\end{enumerate}

The combined object, a folded policy graph keyed by signature with UCB1-selected action statistics, is auditable: an operator can open it after learning and see exactly which actions were selected for each regime, with visit counts, mean rewards, and failure rates. This auditability is itself a deliberate design property; \S\ref{subsec:robustness} returns to its role in reward-signal robustness.

\subsection{Six Layers}
\label{subsec:layers}

AgensFlow is structured into six dependency layers, each independently useful. Table~\ref{tab:layers} summarizes the layer responsibilities and dependency direction.

\begin{table}[H]
\centering
\small
\begin{tabular}{>{\bfseries}r p{0.78\linewidth}}
\toprule
Layer & Responsibility \\
\midrule
5 & \textbf{Pre-flight + governance + reports.} Catch broken
infrastructure (bad API key, exhausted quota, rate-limited
provider) before LLM tokens are spent; halt runs on policy
violations; emit structured per-run artifacts. \\
4 & \textbf{Reward + RelativeJudge.} Compute the signal that flows
into value estimates: rubric-anchored relative-quality score
combined with operational penalties (cost, retries). \\
3 & \textbf{Policy graph + router + LangGraph integration} Folded (signature, action) value table;
reliability-aware UCB1 selection; pure-function routing; LangGraph
dispatch topology. \\
2 & \textbf{Agents + transport + trace.} OpenRouter + Instructor-typed I/O  for
every LLM call; per-skill model bindings; in-memory event
accumulation. \\
1 & \textbf{Schema + regimes + skill registry.} Structured
handoffs, regime detection, activation planning, skill cards
(SKILL.md). \\
0 & \textbf{Web search wrappers.} First-class tool actions (Exa,
Tavily) with provider-aware retry/backoff/clamp semantics. \\
\bottomrule
\end{tabular}
\caption{The six dependency layers of the AgensFlow framework.
  Upper layers depend on lower ones; at runtime all six are active
  simultaneously.}
\label{tab:layers}
\end{table}

The layering is structural, not procedural: at runtime every layer is active simultaneously. The number indicates dependency direction: upper layers depend on lower ones, never the reverse. Layer 0 is numbered lowest because stable tool semantics are a dependency of the runtime, not because retrieval is conceptually prior to coordination. These structural dependencies define the substrate evaluated in the experiments below.

\subsection{Runtime Data Flow}

A single task flows through the substrate as follows:

\begin{enumerate}[leftmargin=*, itemsep=0.2em, topsep=0.3em]
  \item Pre-flight checks gate the run (Layer 5).
  \item Task features are computed; the default rule-based
    \texttt{detect\_regime} function produces a regime label; the
    regime, handoff mask, and bucketed beliefs are folded into a
    signature tuple (Layer 1).
  \item At each routing step, the reliability-aware UCB1 rule in the
    policy graph selects an action from the legal set, invoking a
    cell (planner, memory, web\_search, solver variant, verifier,
    evaluator) or \texttt{skip:X} (Layer 3).
  \item The selected action executes; the trace collector records
    a structured event (Layer 2).
  \item The handoff and belief state are updated, the signature is
    recomputed, and routing returns to the previous step until
    \texttt{terminate} is selected or governance halts the run.
  \item Governance observes the trace and halts the run if a policy
    violation occurs (Layer 5).
  \item When the trajectory completes, \texttt{RelativeJudge} scores it
    relative to same-class peers; the hybrid reward composes
    quality with cost and retry penalties; the (signature, action)
    edges are backed up into the policy graph (Layer 4 + Layer 3).
  \item A structured RunReport is emitted (Layer 5).
\end{enumerate}

The operational form of the hybrid reward is the same as
Equation~\ref{eq:hybrid-reward}, written here with runtime variable
names:
\begin{equation}
  r = w_{\text{quality}} \cdot \text{RJ}
    - w_{\text{cost}} \cdot \frac{\text{tokens}}{\text{cap}}
    - w_{\text{retry}} \cdot \text{retries}
\label{eq:reward}
\end{equation}
where $\text{RJ} \in [0, 1]$ is the \texttt{RelativeJudge} score and the
token term is normalized by the configured cap before weighting.

\subsection{Advantages of AgensFlow}
\label{sec:advantages}

The substrate design gives AgensFlow several advantages over static multi-agent pipelines and one-off model selection sweeps.

\begin{itemize}[leftmargin=*, itemsep=0.3em, topsep=0.3em]
  \item \textbf{Use-case-conditioned coordination.} Routing
    decisions are conditioned on folded task signatures, so the
    system can learn different skill, model, and topology choices
    for recurring task regimes instead of applying one workflow to
    all requests.
  \item \textbf{Topology as a policy variable.} Optional cells can be
    skipped through \texttt{skip:X} actions. This makes the shape of
    a trajectory learnable.
  \item \textbf{Inspectable skill and model provenance.} The policy
    graph records which skill protocols and model-role bindings were
    selected for each signature, with visit counts, reward estimates,
    token statistics, and failure information.
  \item \textbf{Auditable reward.} \texttt{RelativeJudge}, cross-judge
    averaging, per-axis scores, and confidence weighting make the
    reward signal observable rather than treating LLM evaluation as a
    black-box scalar.
  \item \textbf{Reusable coordination priors.} Saved policy graphs
    can warm-start related domains, reducing exploration cost while
    preserving the ability to adapt from new trajectories.
\end{itemize}

\section{Reward and Audit}
\label{sec:reward-audit}

The policy graph is only as useful as the reward signal it learns from. AgensFlow therefore treats evaluation as part of the system rather than as an external reporting step. The reward layer has two
jobs: provide a scalar signal for online backup, and expose enough audit information to detect when the signal itself is biased or unstable.

\subsection{RelativeJudge}

For each completed trajectory $t$ in a group of $N$ same-class peers, an LLM judge ranks the $N$ trajectories side-by-side against an explicit rubric and emits $N$ scalar scores in $[0, 1]$ plus
per-axis sub-scores. The relative-ranking design is used to reduce reward-hacking pressure, a known failure mode in reinforcement learning where the policy learns to exploit weaknesses in the reward signal rather than improve the intended task behavior. Instead of rewarding an internal completion flag or a single absolute score, \texttt{RelativeJudge} compares trajectories against same-class peers under a human-readable rubric. This keeps the reward criterion externally inspectable and editable, while making it harder for the policy to benefit from satisfying a narrow internal signal alone.

The module is a custom implementation inspired by RULER-style relative evaluation \cite{ruler2025}, integrated with the framework's typed agent transport through Instructor and OpenRouter. The framework extends the basic pattern with:

\begin{itemize}[leftmargin=*, itemsep=0.3em, topsep=0.3em]
  \item \textbf{Cross-judge averaging.} The
    \texttt{cross\_judge\_models} configuration accepts a list of
    judges from different model providers (e.g. Anthropic + OpenAI +
    Qwen). The substrate runs each judge independently and averages
    the per-trajectory scores, with per-judge disagreement
    surfaced as confidence telemetry. This mitigates single-judge
    family bias.
  \item \textbf{Per-axis decomposition.} The default rubric has
    four named axes (\texttt{goal\_achievement},
    \texttt{grounding}, \texttt{coordination},
    \texttt{recovery}). \texttt{axis\_weights} composes the scalar
    score as a weighted sum.
  \item \textbf{Confidence weighting.} Reward backups are
    multiplied by judge confidence (computed from disagreement
    std), so noisy reward observations contribute less to value
    estimates.
\end{itemize}

\section{Empirical Findings}
\label{sec:findings}

This section reports cross-domain validation on two 60-task evaluation corpora. The goal is to test whether the same coordination substrate can operate across different topic domains, and whether a policy graph learned on one domain can provide a useful warm-start prior for another. The framework, action surface, reward interface, and coordination formulation are held fixed across both evaluations.

\subsection{Evaluation Corpora}
\label{subsec:evaluation-datasets}

The first corpus contains 60 distributed-systems incident tasks. Tasks cover distributed-systems concepts and incident reasoning patterns such as Paxos and Raft, logical clocks, gossip, CRDTs, consistent hashing,
failure detection, and multi-evidence synthesis. This corpus is used to produce the policy graph for warm-start initialization in the transfer arm reported below. The final graph contains 443 nodes.

The second corpus contains 60 synthetic security-advisory tasks spanning six vendor pairs. It is designed to test cross-document reasoning, mitigation correctness, evidence synthesis, and out-of-corpus failure
modes. The task pool covers eight scenario classes: C1 procedural; C2 single-doc; C3 cross-document multi-vendor; C4 synthesis; C5 out-of-corpus ambiguous; C6 procedural-derivative; C7 mitigation correctness; and C8 cross-vendor pair.

The two corpora differ in topic domain but share the same observable feature taxonomy, folded-signature construction, action surface, and reward interface. This makes the synthetic security-advisory corpus a transfer test for learned coordination priors. 

\subsection{Experimental Configuration}

\textbf{Substrate.} The skill-variant pool is held fixed across both domains: 9 (skill protocol $\times$ model) solver cells, a planner, memory, two web-search providers, two verifier variants, and an evaluator. The \texttt{skip:X} action is enabled.

\textbf{Reward.} The live UCB1 signal uses the hybrid reward in
Equation~\ref{eq:reward}, with Claude Haiku 4.5 as
the single judge.

\textbf{Compared arms.} Four arms are evaluated on the same 60 security-advisory tasks: (i) \emph{baseline}: fixed 7-cell pipeline, single pass, no learning; (ii) \emph{no-skip ablation}: 8 epochs, variant-pool learning with \texttt{skip:X} forced off; (iii) \emph{main}: 8 epochs, \texttt{skip:X} enabled, cold-start; and (iv) \emph{warm-start}: 8 epochs, \texttt{skip:X} enabled, initialized from the 443-node distributed-systems policy graph.

\textbf{Audit.} Final trajectories from all four arms are re-scored post-hoc under a cross-family three-judge ensemble (Claude Haiku 4.5 + GPT-5.4 mini + Qwen3.6-flash) with 100\% axis-population compliance.

\begin{figure}[H]
  \centering
  \includegraphics[width=\linewidth]{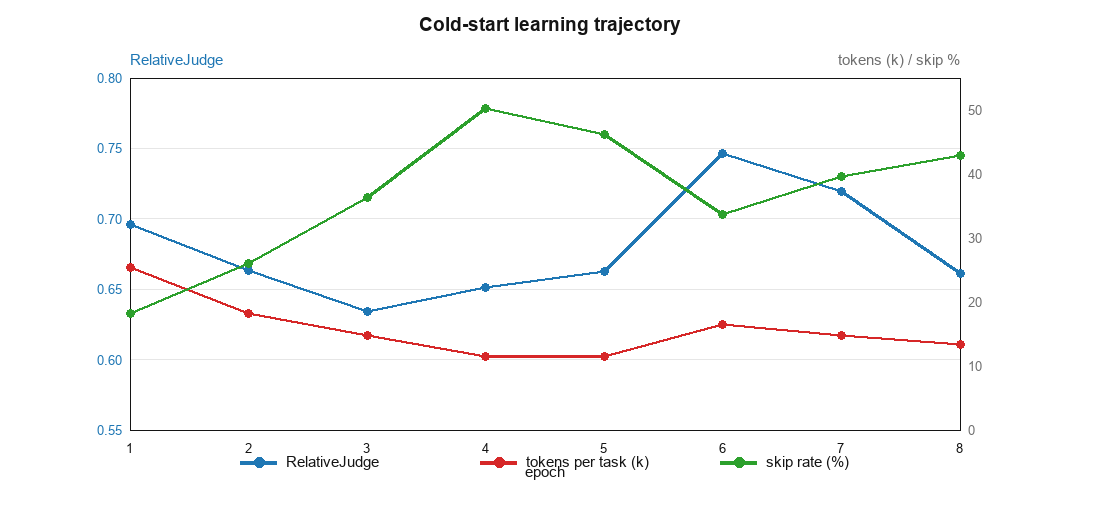}
  \caption{\textbf{Cold-start learning trajectory.} The main substrate initially compresses cost as \texttt{skip:X} usage rises,
    then rebalances after quality drops. The pattern supports that skip decisions are being adjusted by reward feedback rather than fixed by the initial pipeline.}
  \label{fig:e09-learning-skip}
\end{figure}

Figure~\ref{fig:e09-learning-skip} shows the cold-start learning trajectory for the main run before the aggregate results are reported.

\subsection{Cross-Domain Transfer}
\label{subsec:cross-domain}

Table~\ref{tab:e09-arm-summary} reports the arm-level comparison at
converged plateau. The main substrate beats the fixed baseline under
cross-family audit at epochs 6 to 8.

\begin{table}[H]
\centering
\caption{\textbf{Arm-level comparison.} Learning arms are reported
  at plateau epochs 6 to 8; the fixed baseline is a single-pass run on
  the same 60 tasks. RJ (sj) denotes \texttt{RelativeJudge} scores from the single judge, while RJ (3j) denotes the cross-family three-judge audit score.}
\label{tab:e09-arm-summary}
\vspace{0.3em}
\small
\begin{center}
\begin{tabular}{l r r r r}
\toprule
arm & RJ (sj) & RJ (3j) & tokens & $\Delta$ vs baseline (3j) \\
\midrule
fixed baseline           & 0.622 & 0.749 & 12{,}960 & n/a \\
no-skip ablation         & 0.662 & n/a   & 25{,}198 & token: $+94\%$ \\
\textbf{main (plateau)}  & \textbf{0.709} & \textbf{0.817} & \textbf{14{,}870} & RJ $+0.068$ / tok $+15\%$ \\
warm-start (plateau)     & 0.761 & 0.829 & 13{,}371 & RJ $+0.080$ / tok $+3\%$ \\
\bottomrule
\end{tabular}
\end{center}
\end{table}

Table~\ref{tab:e09-per-class} breaks the audited plateau result down by scenario class. The same deltas were previewed in Figure~\ref{fig:e09-per-class-lift}.

\begin{table}[H]
\centering
\caption{\textbf{Per-class plateau comparison under 3-judge audit.}
  Values compare the main substrate at plateau against the fixed
  baseline.}
\label{tab:e09-per-class}
\vspace{0.3em}
\small
\begin{center}
\begin{tabular}{l r r r}
\toprule
class & baseline & main & $\Delta$ \\
\midrule
C1 procedural               & 0.848 & 0.806 & $-0.042$ \\
C2 single-doc               & 0.758 & 0.847 & $+0.089$ \\
\textbf{C3 cross-doc multi-vendor} & 0.675 & \textbf{0.857} & $\mathbf{+0.181}$ \\
C4 synthesis                & 0.776 & 0.827 & $+0.050$ \\
C5 out-of-corpus            & 0.802 & 0.778 & $-0.024$ \\
C6 procedural-derivative    & 0.794 & 0.798 & $+0.004$ \\
\textbf{C7 mitigation correctness} & 0.658 & \textbf{0.790} & $\mathbf{+0.131}$ \\
\textbf{C8 cross-vendor pair}      & 0.673 & \textbf{0.829} & $\mathbf{+0.156}$ \\
\bottomrule
\end{tabular}
\end{center}
\end{table}

The substrate wins on 5 of 8 classes under audit, ties on C6, and trades narrowly on C1 and C5. The gains concentrate on coordination-heavy classes: cross-document multi-vendor reasoning (C3, $+0.18$), cross-vendor pair analysis (C8, $+0.16$), and mitigation correctness (C7, $+0.13$). These are the classes where multi-step synthesis and selective verification matter most. Procedural classes such as C1 and C6, which can be solved adequately by running every cell every time, are statistical ties.

The evaluated default solver is \texttt{solver\_cot\_haiku}. The substrate moved off the default on 7 of 8 classes, selecting evidence-template variants on C1 and C2, concise-template variants on C3, C4, C5, C6, and C8, and keeping the default only on C7, where the cost of an incorrect answer is highest. This pattern suggests that the learned policy is not merely selecting cheaper variants uniformly, but adapting routing decisions to the structure and risk profile of each task class.

\subsection{Policy Transfer via Warm-Start}
\label{subsec:warm-start}

\begin{figure}[H]
  \centering
  \includegraphics[width=\linewidth]{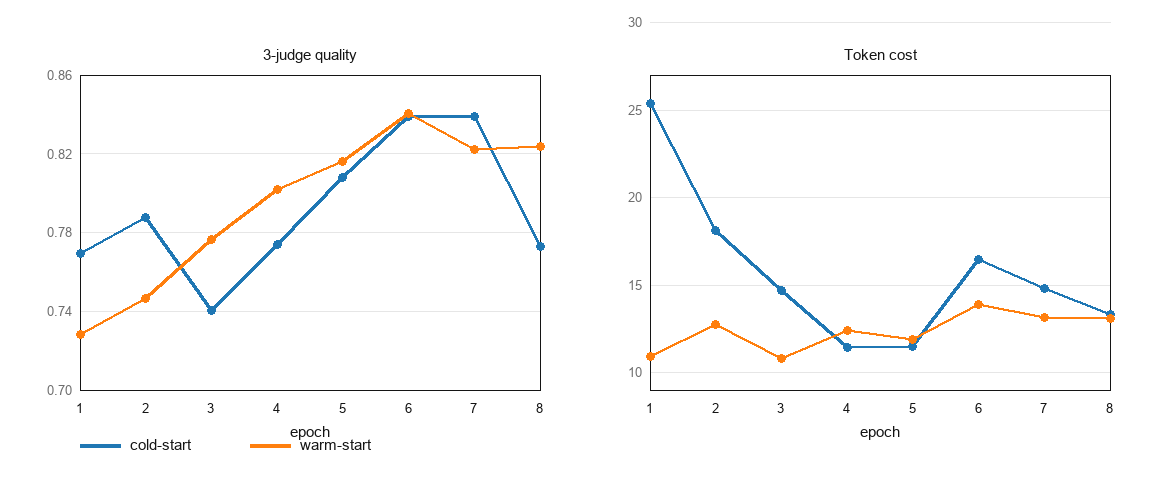}
\caption{\textbf{Warm-start transfer.} Warm-starting from the
  distributed-systems policy graph reduces early exploration cost on the
  synthetic security-advisory corpus while preserving plateau quality
  under cross-family audit. The figure also shows why the single-judge
  result requires audit: quality differences are modest under 3-judge
  scoring, while token compression is judge-independent.}
  \label{fig:e09-warm-transfer}
\end{figure}

The warm-start arm initializes with the 443-node policy graph learned on the distributed-systems incident corpus, which is structurally similar but topically different from the security-advisory corpus evaluated here. The single-judge data shows warm-start beating cold-start on
\texttt{RelativeJudge} in 8 of 8 epochs ($+0.052$ plateau, $+0.055$ full-run mean). The cross-family audit shifts this picture substantially, as shown numerically in Table~\ref{tab:e09-warm-start} and visually in Figure~\ref{fig:e09-warm-transfer}.

\begin{table}[H]
\centering
\caption{\textbf{Warm-start transfer under 3-judge audit.} Warm-starting
  from the distributed-systems policy graph reduces exploration cost,
  while audited quality shows a small plateau gain rather than the larger
  improvement suggested by the single-judge signal.}
\label{tab:e09-warm-start}
\vspace{0.3em}
\small
\begin{center}
\begin{tabular}{r r r r}
\toprule
epoch & cold (3j) & warm (3j) & $\Delta$ \\
\midrule
1 & 0.769 & 0.728 & $-0.041$ \\
2 & 0.788 & 0.747 & $-0.041$ \\
3 & 0.740 & 0.776 & $+0.036$ \\
4 & 0.774 & 0.802 & $+0.027$ \\
5 & 0.808 & 0.816 & $+0.008$ \\
6 & 0.839 & 0.841 & $+0.001$ \\
7 & 0.839 & 0.822 & $-0.017$ \\
8 & 0.773 & 0.824 & $+0.051$ \\
\midrule
\emph{plateau (6 to 8)} & 0.817 & 0.829 & \textbf{+0.012} \\
\emph{full-run mean}  & 0.791 & 0.794 & $+0.003$ \\
\bottomrule
\end{tabular}
\end{center}
\end{table}

Warm-starting from the distributed-systems policy graph reduces exploration cost on the security-advisory corpus while preserving audited plateau quality. Under the single-judge signal, warm-start shows a quality advantage over cold-start across all epochs, with $+0.052$ plateau lift and $+0.055$ full-run mean lift. The cross-family audit shows that most of this apparent quality advantage is judge-sensitive: the audited plateau lift is $+0.012$, and the audited full-run lift is $+0.003$. The cost effect is stronger and stable across judges: warm-start uses ${\sim}10\%$ fewer tokens at plateau and ${\sim}21\%$ fewer tokens across the full run. It also preserves targeted audited quality gains on C5 ($+0.081$) and C7 ($+0.049$). These results support that learned coordination priors can reduce exploration cost without degrading plateau quality, while also showing why reward audit belongs inside the evaluation protocol.

\subsection{Reward-Signal Robustness as a First-Class Concern}
\label{subsec:robustness}

The cross-family audit shows that reward choice can affect headline interpretation. Under the single-judge signal, warm-start appears to provide a larger quality advantage over cold-start than it does under the three-judge audit. This does not change the main transfer result: warm-start reduces exploration cost and preserves plateau quality. It does, however, show that reward signals should be treated as part of the coordination substrate rather than as a neutral measurement layer.

The implication extends beyond reward modeling. If even a structured peer-relative judge can shift conclusions about agentic trajectories, then hand-designing coordination policies over interacting task, skill, model, and topology choices is intrinsically fragile. A one-shot topology choice made from engineering intuition has less structure than the judge: no repeated comparison, no explicit axis decomposition, and no reward audit. The $(\text{task} \times \text{skill} \times \text{model} \times \text{topology})$ interaction surface is therefore too large and too noisy to navigate reliably by intuition alone.

A similar issue appears on the control side. Adding retries to a fixed pipeline can improve per-step success probability, but each retry is an independent call and does not sharpen the policy. With base per-step success $p$ and $n$ retries on a $k$-stage pipeline, end-to-end success is $(1 - (1-p)^n)^k$; improving reliability this way can require many additional calls without producing reusable coordination knowledge. AgensFlow instead records which routing choices worked, under which signatures, and backs those observations up into the policy graph.

This observation directly informs the framework's treatment of reward-signal robustness. AgensFlow uses a cross-judge setup to compare how reward signals affect policy conclusions across model families. Routing decisions remain auditable at the signature level, and the hybrid-reward weights can be adjusted to reflect different operational priorities. Reward auditing is therefore part of the coordination substrate itself.

\section{Discussion}
\label{sec:discussion}

The empirical result is not only that one experimental arm scored higher than another. The deeper finding is that the coordination surface is too large, too regime-dependent, and too judge-sensitive to navigate by intuition alone. Skills, models, retrieval, verification, skipping, reward design, and cost constraints interact. A choice that is beneficial for one task signature can be wasteful or brittle for another. This is why the central object in AgensFlow is not a fixed workflow, but an auditable routing policy over repeated task regimes.

For production multi-agent systems, the relevant question is therefore not "which model is best?" but "which coordination policy works under this observable task regime and operating constraint?" Fixed pipelines hide this question inside hand-written orchestration. AgensFlow makes the question explicit: it logs trajectories, aggregates repeated reward observations, exposes skill/model/topology choices, and keeps reward audit attached to the learning substrate. However, the framework does not replace product judgment. It makes the routing surface observable and improvable from reward while acknowledging that the reward signal itself must be audited.

\section{Limitations}
\label{sec:future}

The current substrate exercises a linear-with-skip topology with single-judge live reward. The evaluation covers this topology class only. Alternative coordination topologies, including parallel or swarm-style execution, branching coalitions with verifier-mediated merge, and hierarchical decomposition with sub-agent planners, are part of the framework's design surface but were not exercised in the reported experiments. Transfer was tested across two corpora that share a common feature taxonomy. Future work should test transfer under different feature taxonomies, evaluate alternative planning regimes over the broader topology classes above, and incorporate live cross-judge reward. Reward-signal effects surfaced by the cross-family audit also warrant further study under different deployment conditions.

\section{Conclusion}
\label{sec:conclusion}

AgensFlow demonstrates that coordination policy for multi-agent systems can be learned online from repeated reward signals over a four-way interaction surface of task signatures, skills, models, and topology choices. The cross-domain validation reported here shows that the mechanism transfers to a structurally novel corpus, with substrate-versus-baseline quality gains concentrated in the classes where coordination matters most.

The results also show that reward robustness is part of the coordination problem. The cross-family audit changes the magnitude of the warm-start quality advantage while preserving the core finding that learned
coordination priors reduce exploration cost without degrading plateau quality. This motivates reward auditability, including cross-judge configuration, as a primary design feature rather than an external evaluation step.

The framework is released open-source. The experiments and policy graphs learned during the reported evaluations are reproducible end-to-end from the repository.

\printbibliography

\end{document}